\documentclass[journal=jctcce,manuscript=article, layout=twocolumn]{achemso}
\usepackage[T1]{fontenc}       
\usepackage                             {subcaption}
\usepackage{afterpage}
\usepackage{amsmath}
\usepackage{pdfpages}

\newcommand{\bs}[1]{\mathrm{\mathbf{#1}}}
\newcommand{\R}{\bs{r}}
\newcommand{\W}{\bs{w}}
\newcommand{\der}[0]{\mathrm{d}}

\usepackage{xcolor}

\author{Martin Reinhardt}
\affiliation[HITS]
{Molecular and Cellular Modeling Group, Heidelberg Institute for Theoretical
Studies (HITS), Schlo\ss-Wolfsbrunnenweg 35, 69118 Heidelberg, Germany}
\alsoaffiliation[MPI]
{Department of Theoretical and Computational Biophysics, Max Planck Institute for Biophysical Chemistry, Am Fassberg 11, 37077 G\"ottingen, Germany}
\author{Neil J. Bruce}
\affiliation[HITS]{Molecular and Cellular Modeling Group, Heidelberg Institute for Theoretical Studies (HITS), Schlo\ss-Wolfsbrunnenweg 35, 69118 Heidelberg, Germany}

\author{Daria B. Kokh}
\affiliation[HITS]{Molecular and Cellular Modeling Group, Heidelberg Institute for Theoretical Studies (HITS), Schlo\ss-Wolfsbrunnenweg 35, 69118 Heidelberg, Germany}

\author{Rebecca C. Wade}
\affiliation[HITS]{Molecular and Cellular Modeling Group, Heidelberg Institute for Theoretical Studies (HITS), Schlo\ss-Wolfsbrunnenweg 35, 69118 Heidelberg, Germany}
\alsoaffiliation{Center for Molecular Biology (ZMBH), University of Heidelberg, Im Neuenheimer Feld 282, 69120 Heidelberg, Germany}
\alsoaffiliation{Interdisciplinary Center for Scientific Computing (IWR), Heidelberg University, Im Neuenheimer Feld 368, 69120 Heidelberg, Germany}
\email{rebecca.wade@h-its.org}
\phone{+49-6221-533 247}
\fax{+49-6221-533 298}

\title[Brownian Dynamics Simulations With Surfaces]
  {Brownian Dynamics Simulations of Proteins in the Presence of Surfaces: Long-range Electrostatics and Mean-field Hydrodynamics}

\abbreviations{HI,DH,BD}

\begin{document}

	\begin{tocentry}	
	\centering
	\includegraphics[width=\linewidth]{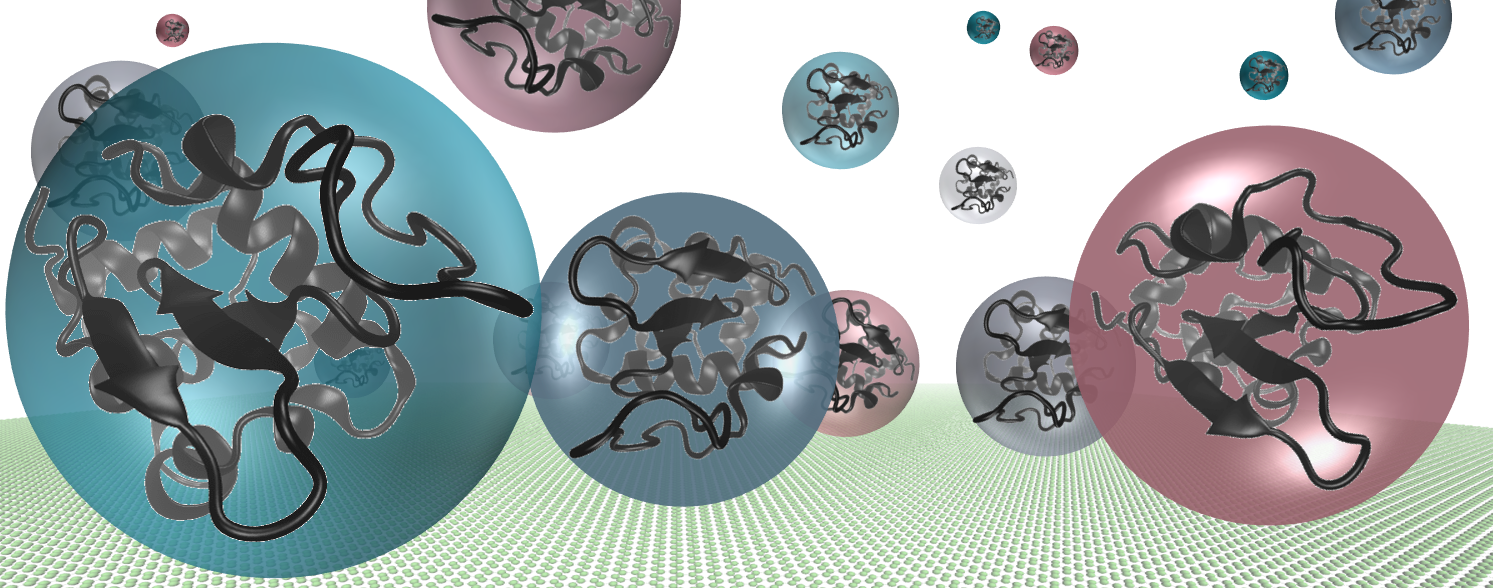}

\end{tocentry}

\begin{abstract}
Simulations of macromolecular diffusion and adsorption in confined environments can offer valuable mechanistic insights into numerous biophysical processes. In order to model solutes at atomic detail on  relevant time scales, Brownian Dynamics simulations can be carried out with the approximation of rigid body solutes moving through a continuum solvent. This allows the precomputation of interaction potential grids for the solutes, thereby allowing the computationally efficient calculation of forces. However, hydrodynamic and long-range electrostatic interactions cannot be fully treated with grid-based approaches alone. Here, we develop a treatment of both hydrodynamic and electrostatic interactions to include the presence of surfaces by modeling grid-based and long-range interactions. 
 We describe its application to simulate the self-association and many-molecule adsorption of the well-characterized protein Hen Egg-White Lysozyme to mica-like and silica-like surfaces.  We find that the computational model can recover a number of experimental observables of the adsorption process and provide insights into their determinants.  The computational model is implemented in the Simulation of Diffusional Association (SDA) software package.  
 
\end{abstract}

\section{Introduction}

The diffusion of macromolecules, such as proteins, is essential for cellular processes, including signal transmission and the transport of matter\cite{Mika2011, Mourao2014, Luby-Phelps2013}. Diffusion-related properties are often highly dependent on the environment of the molecules\cite{Minton2001, Rivas2004, Minton2006, Klumpp2013, Kuznetsova2015, Rivas2016}, with numerous biological consequences \cite{Zhou2013, Guin2019}. Furthermore, interactions with surfaces influence the diffusive behavior of these molecules. Firstly, in the case of attractive interactions, adsorption processes occur, including protein aggregation at confining\cite{Vieira2009} or artificial surfaces such as implants \cite{Wang2002}, and these are of interest for their implications for drug delivery and biosensing \cite{Gray2004}. Secondly,  inside  cells, surfaces restrict diffusive motion, e.g., due to the plasma membrane surrounding the cellular volume as a whole, or because of internal cytoskeletal elements. These restrictions have been identified as a major determinant of macromolecular kinetics and reactivity, and have been analyzed theoretically \cite{Minton1992, Minton1995, Minton2001}, as well as experimentally \cite{Zhou2008, Hoffmann2009, Konorty2009}. Predicting the effects of such surfaces on protein diffusion is, however, a highly complex task \cite{Dix2008, Cohavi2010, Ozboyaci2016b}.

A number of factors contribute to the diffusive behavior of molecules in the presence of surfaces. Aside from direct steric obstruction, hydrodynamic interactions (HI) with other molecules \cite{Ando2013} and with surfaces have been shown to lead to a significant reduction in macromolecular self-diffusion \cite{Dufresne2000, Doster2007, Lisicki2016, Czajka2019}. In addition, occurrences of non-Gaussian mean-squared displacements near surfaces have been observed for colloidal solutions\cite{Skaug2013, Matse2017, Czajka2019}. The adsorption processes of molecules to surfaces are affected by long-range electrostatic interactions and, therefore, depend on the charge density of the given surface and the surrounding salt concentration\cite{Antosiewicz2020}. Furthermore, short-range interactions, such as van-der-Waals forces, also affect the orientation and structure of molecules on surfaces~\cite{Buijs1997, Ozboyaci2016, Romanowska2015}. 

Computational approaches can provide mechanistic insights for systems that are often hard to access experimentally with techniques such as nuclear magnetic resonance \cite{Li2013} or fluorescence spectroscopy \cite{Ignatova2004}. However, while molecular dynamics (MD) simulations of dense protein solutions in atomic detail with explicit solvent models have been performed in recent years \cite{Yu2016, Feig2017, VonBulow2019}, they require very large computational resources, and the short lengths of the simulated timescales often make it hard to obtain statistically significant results on diffusion-related processes. Therefore, it is reasonable to employ different levels of detail in computer simulations, depending on the length and time scales of the system of interest. While MD simulations are suitable for capturing short-range macromolecule-surface interactions \cite{Wei2012, Kubiak2016}, Brownian Dynamics (BD) simulations can be used to study the properties defining processes occurring on longer time scales, such as the the kinetic and structural properties of macromolecular adsorption\cite{Brancolini2012, Brancolini2015}. BD simulations have been performed of spherical particle models of the macromolecules with a charged surface\cite{Ravichandran2000, Carlsson2004, Gorba2004}. So far, however, BD simulations with structurally detailed molecular models have mostly been restricted to systems consisting of one solute and a surface, although they have been used to simulate the diffusion of many hydrophobin molecules to an uncharged surface\cite{Mereghetti2011} and a few diffusing hen egg white lysozyme (HEWL) molecules to a charged surface \cite{Ravichandran2001}. 

The Simulation of Diffusional Association (SDA) BD software package\cite{Gabdoulline1997, Gabdoulline1998, Martinez2015} can be used to simulate the motion of pairs of macromolecules, of (dense) solutions of macromolecules, and of macromolecules in the presence of surfaces.  SDA uses a rigid-body model, that permits intermolecular interaction forces to be computed efficiently by precomputation of their interaction potentials on three-dimensional discretized grids while at the same time calculating the interactions on an atomic basis rather than approximating the solute by, e.g., a sphere or an ellipsoid. HI between solutes is modelled by a mean-field approximation described by \citet{Mereghetti2012}.  Here, we first extend this approach to introduce a treatment of solute-surface HI by a first-order approximation that can be used to simulate the adsorption of either a single solute or of multiple solutes to a surface. Then, for long-range electrostatic interactions that extend beyond the dimensions of the electrostatic potential grids, we introduce a Debye-H\"uckel (DH) approximation for surfaces. This treatment complements the DH approximation that was previously introduced for long-range electrostatic interactions between solutes \cite{Mereghetti2014}, for which we here describe an improved treatment of the transition between grid and DH formulations at the grid edges.
We describe the validation and application of these new computational models to test systems containing the experimentally well-characterized protein, HEWL. We first compute and analyze the rate of approach of two HEWL proteins, and then we simulate the adsorption process of over a hundred HEWL molecules to attractive mica- and silica-like surfaces for different bulk protein concentrations. These simulations allow us to investigate the contributions of HI and long-range electrostatic interactions to these protein-surface adsorption processes. 

\section{Theory and Methods}

\subsection{Brownian Dynamics Simulations}

In BD simulations,  a mesoscopic model is employed to simulate the motion of solutes over length and time scales relevant for Brownian motion, that is the random motion of solutes in fluids where the solutes move much more slowly than the  solvent molecules. An implicit solvent model is used: besides the systematic forces, the effect of the solvent is included through stochastic sampling of the collisions with solvent molecules. In SDA, solute trajectories are propagated according to the algorithm described by Ermak and McCammon \cite{Ermak1978}. The trajectories are composed of successive displacements of the solutes, each taken over a short time step $\Delta t$. The translational displacement of a Brownian particle $i$ is calculated as
\begin{align}
\begin{aligned}
\R_i^{n+1} = \R_i^n &+ \sum_j \frac{\partial \bs{D}_{ij}}{\partial \R_j^n}\:\Delta t\\
&+ \sum_j \frac{\bs{D}_{ij}}{k_\textnormal{B}T}\: \bs{F}_{i}^n \:\Delta t + \bs{R}_i \; 
\label{eq:ermak-mccammon}
\end{aligned}    
\end{align}
where $\R_i^n$ denotes the position of the center of geometry of particle $i$. The superscript $n$ indicates that the variable is to be evaluated at the beginning of the time step, $n+1$ is the result after the time step. $\bs{F}_{i}$ is the total systematic force acting on particle $i$ and $\bs{D}_{ij}$ is a $3 \times 3$ subtensor of the hydrodynamically-coupled diffusion tensor $\bs{D}$ of the system of Brownian particles, where the diagonal subtensor $\bs{D}_{ii}$ is the infinite dilution diffusion tensor of particle $i$ and the off-diagonal subtensors account for the configuration-dependent hydrodynamic coupling between particles $i$ and $j$.  $\bs{R}_i$ is a stochastic displacement vector that takes into account the collisions with the solvent molecules. It is drawn from a Gaussian distribution with mean $<\bs{R}_i> = 0$ and covariance $<\bs{R}_i\bs{R}_j> = 2\bs{D}_{ij}\Delta t$ for all $i$ and $j$. 

Propagating a system of Brownian solutes that each consist of a large number of particles using Eq. \ref{eq:ermak-mccammon} is computationally expensive, as the calculation of the hydrodynamically-coupled stochastic term of the BD propagation step requires Cholesky factorization of the tensor $\bs{D}$  at every time step, which scales as $\mathcal{O}(N^3)$ for $N$ Brownian solutes, although with approximations this can be reduced to $\mathcal{O}(N^{2.25})$\cite{Schmidt2011, Saadat2014} or $\mathcal{O}(N^{2})$\cite{Geyer2009}. This compares with $\mathcal{O}(N^{2})$ for the calculation of the interparticle systematic forces, which can be reduced to 
$\mathcal{O}(N)$ through the use of distance cutoffs.

As SDA was initially developed to simulate the association of a pair of solutes,\cite{Gabdoulline1997} approximations were made to simplify Eq. \ref{eq:ermak-mccammon}. It was assumed that each solute diffuses as a single spherical hydrodynamic bead and that hydrodynamic interactions are negligible, due to the low solute concentration. This meant that all off-diagonal terms in Eq.~\ref{eq:ermak-mccammon} could be ignored and the diagonal subtensors could be could be replaced with scalar isotropic diffusion translational coefficients $D^\textnormal{t}_i$, resulting in :
\begin{equation}
\R_i^{n+1} = \R_i^n + \frac{D^\textnormal{t}_i}{k_\textnormal{B}T} \: \bs{F}_i^n \:\Delta t + \bs{R}_i \;
\label{eq:step_const_diff}
\end{equation}
Importantly, the stochastic vector $\bs{R}_i$ is no longer configuration-dependent and can be drawn from a precomputed distribution. The propagation of the rotation is calculated through an analogous equation :
\begin{align}
\begin{aligned}
\W_i^{n+1} = \W_i^n + \frac{D_i^\textnormal{r}}{k_\textnormal{B}T} \:\bs{T}_i^n \:\Delta t + \bs{W}_i 
\end{aligned}    
\end{align}
where $\W_i$ and $\bs{W}_i$ describe the orientation and the stochastic rotation vector, respectively, of solute $i$. $D_i^\textnormal{r}$ denotes the rotational diffusion coefficient, and $\bs{T}_i^n$ describes the sum of torques acting on solute $i$. 

\subsection{Mean-Field Intersolute Hydrodynamic Interactions}
\label{subsec:hi}

As SDA was extended to allow the simulation of protein solutions\cite{Mereghetti2010}, it became apparent that the assumption that intersolute hydrodynamic interactions could be ignored was no longer valid for increasing concentration of solutes. Therefore a mean-field hydrodynamic model\cite{Mereghetti2012, Tokuyama1994, Tokuyama2011} was developed in which $D_{i}^\textnormal{t}$ is replaced by a local occupied volume fraction-dependent diffusion coefficient  $D_i(V_i^\textnormal{frac})$ and the translational displacement equation becomes \cite{Mereghetti2012}:

\begin{equation}
\R_i^{n+1} = \R_i^n +\frac{D_i(V_i^\textnormal{frac})}{k_\textnormal{B}T}\: \bs{F}_i^n \:\Delta t + \bs{R}_i 
\label{eq:step_vfrac_diff}
\end{equation}

where $V_i^\textnormal{frac} = \sum_j \nu_j \slash V_i$ denotes the dimensionless local occupied volume fraction around solute $i$. It is obtained by summing over the volumes $\nu_j$ of the surrounding solutes $j$, calculated by approximating these as spheres of radius $a_j$. The sum only includes the solutes within a sphere with radius $R_\textnormal{cut}$ with volume $V_i=(4\pi \slash 3)\: (R_\textnormal{cut})^3$ centered on the solute $i$. The volume fraction dependent short-time translational diffusion coefficient $D_i(V_i^\textnormal{frac})$ is then obtained using the Tokuyama model \cite{Tokuyama1994, Tokuyama2011}, derived for a concentrated hard-sphere suspension of solutes interacting with both direct and hydrodynamic interactions. An equation analogous to Eq.~\ref{eq:step_vfrac_diff} is used for the rotational motion with the volume fraction dependent short-time rotational diffusion coefficient obtained by using the model derived by \citet{Cichocki2000}, which includes the lubrication forces as well as two- and three-body expansions of the mobility functions. For a larger number of solutes, the approach correctly reproduces the short and long-time diffusion coefficients\cite{Mereghetti2012} while avoiding the computationally expensive Cholesky factorization.

\subsection{Hydrodynamic Interactions in the Presence of a Surface}

When a solute moves in solution, it creates a flow field, i.e., motion of the fluid, which affects the motion and diffusive behavior of surrounding solutes. In the presence of a surface, this flow field is reflected at the surface, thereby giving rise to additional effects on the solutes. While the inter-solute mean-field approach described in the last section allows concentrated solute solutions to be simulated, it is not able to model the hydrodynamic effects of obstructing surfaces. In this work, we extend this model to account for these effects. 

In doing so, we distinguish between two different effects: Firstly, a single solute near a surface is affected by its own reflected flow field, and as a result experiences a large reduction in its diffusion. Secondly, a solute in a crowded environment also interacts with the reflected flow fields from the surrounding solutes. Naturally, these two effects are correlated. However, computing HI correctly up to a high order becomes difficult both analytically and computationally for as few as two\cite{Dufresne2000} or three spheres in the presence of a surface, and, therefore, is often not accounted for in BD simulations at all. In order to estimate the magnitude of the effect of HI on many solutes close to a surface, we combine the above two effects in a linear first-order approach, thereby approximating them as uncoupled. 

The effect of HI on the diffusive behavior of a single sphere in the presence of a planar surface is analytically accessible and, due to direct coupling of its movement with the change of its image flow field, is known to lead to a large reduction in the effective diffusion coefficient close to the surface. Due to the symmetries of the problem, the translational propagation step can be easily split into a step in the plane parallel to the surface and a step in the direction perpendicular to the surface. The relative reduction, $\zeta_\perp$,  in the short-time diffusion coefficients of a sphere perpendicular to the surface has been determined by Lorentz~\cite{Lorentz1907} and Faxen~\cite{Faxen1922} :

\begin{align}
\begin{split}
\zeta_\perp(z) = & \;\frac{9}{16}\frac{a^{HI}}{z} + \frac{1}{8}\left(\frac{a^{HI}}{z}\right)^3 \\
&- \frac{45}{256} \left(\frac{a^{HI}}{z}\right)^4 - \frac{1}{16}\left(\frac{a^{HI}}{z}\right)^5
\end{split}
\label{eq:lorentz_faxen}
\end{align}

where $a^{HI}$ is the sphere's hydrodynamic radius and $z$ the height of the sphere's center above the surface. They also calculated a result for the relative reduction in the diffusion coefficient parallel to the surface ($\zeta_\parallel(z)  =1-\frac{9}{8}\frac{a^{HI}}{z}$) which is, however, only valid at $a^{HI}\slash z < 1\slash 4$, i.e., when the sphere's center is farther away from the surface than four times the sphere's radius. More refined calculations were conducted by Brenner et al.~\cite{Brenner1961, Goldman1967}, leading to the analytical result:

\begin{equation}
\resizebox{.95\hsize}{!}{$\begin{split}
	&\zeta_\parallel(z) = \left[
	\frac{4\sinh\alpha}{3}\sum_{n=0}^\infty
	\frac{n(n+1)}{(2n-1)(2n+3)} \right.\\
	& \left. \left(
	\frac{2\sinh[(2n+1)\alpha]+(2n+1)\sinh[2\alpha]}
	{(2\sinh[(n+1\slash 2)\alpha])^2 - ((2n+1)\sinh\alpha)^2} -1 \right)\right]^{-1}
	\end{split}$} 
\label{eq:brenner}
\end{equation}

where $\alpha = \cosh^{-1}((a^{HI}\slash z)^{-1})$. A numerical summation at every time step in the simulation would be too costly computationally. Therefore, we conduct the numerical summation for $\zeta_\parallel(z)$ in Eq.~\ref{eq:brenner} once in $a^{HI}\slash z$ and determine a third order polynomial fit. The details are described in section~2 of the Supporting Information.  

The solute is now first assumed to diffuse with a short-time diffusion coefficient obtained by the mean-field approach. In the presence of a surface, this coefficient is further lowered by the relative reduction of Eq.~\ref{eq:lorentz_faxen} perpendicular to the surface, 
and the third order approximation of Eq.~\ref{eq:brenner} parallel to the surface. The resulting short-time diffusion coefficient is then used in the propagation step for BD simulations of Eq.~\ref{eq:step_vfrac_diff}. The resulting diffusion coefficient for motion in the plane parallel to the surface is given by
\begin{equation}
D_i^\parallel(V_i^\textnormal{frac}) = D_i(V_i^\textnormal{frac}) \zeta_\parallel(z)  \;
\end{equation}
and, equivalently, the diffusion coefficient for motion in the direction perpendicular to the surface is obtained through multiplying by $\zeta_\perp(z)$. The reduction of the rotational short-time diffusion coefficient due to HI between a single solute and the surface is not included, as it is much smaller than that for the translational short-time diffusion\cite{Cichocki1998} and is only apparent at very small surface-solute separations. The larger reduction of the rotational diffusion of a solute as a result of its crowded environment is, however, accounted for as described above. 

For the second effect -- the interaction of the reflected flow field with surrounding solutes -- we use the method of images from hydrodynamic theory\cite{Ainley2008}. To satisfy the boundary condition of zero flow field perpendicular to a surface, the reflected solute can conceptually be calculated by using solutes with positions mirrored at the surface. Due to the linearity of the Stokes equation, the superposition of the initial flow field and the one from the mirrored solutes yields the final flow field. For a non-zero flow field parallel to the surface, higher order terms arise. For a discussion of the accuracies of the mobility matrices for a single sphere see, for example, reference~\citenum{Cichocki2000}. However, for cases without collective motion parallel to the surface, we restrict ourselves to the force monopole. 
 
\begin{figure}[t!]
	\includegraphics[width=3.33in]{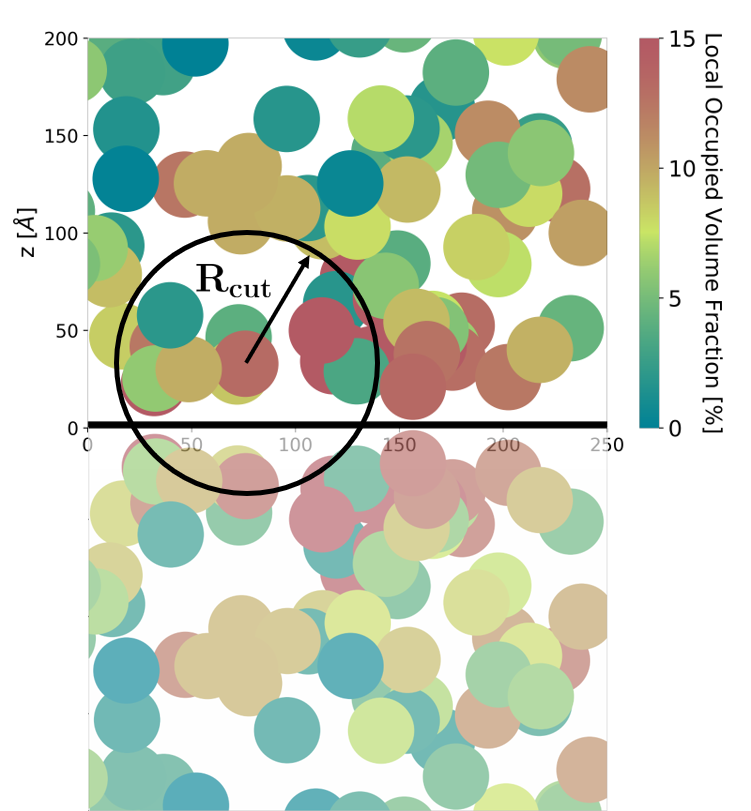}
	\caption{Image solute method for computing HI for solutes diffusing in the presence of a surface. The color indicates the local occupied volume fraction $V_\textnormal{frac}$ within the cutoff sphere with $R_\textnormal{cut} = 4 a^{HI}$ with the solute with radius $a^{HI}$ of interest placed at the center. To resemble HEWL in the adsorption simulation, $a^{HI}$ was set to 15~\AA\ in this figure. The simulated volume is viewed as an orthographic projection viewed along the plane of the surface. The HI of a solute with a surface can be included by considering the interactions with image solutes with mirrored positions on the other side of the surface. The occupied volume fraction is then used with the hydrodynamic mean-field approach.}
	\label{fig:hydro_mirror}
\end{figure}

Using the method of images with this assumption, we extend the mean-field approach beyond the surface by assuming an image flow field created by a mirrored configuration of solutes on the other side of the surface. Figure~\ref{fig:hydro_mirror} shows a snapshot of a simulation with spheres diffusing in the presence of a surface. For each sphere, the local occupied volume fraction is calculated within a cutoff sphere with $R_\textnormal{cut}$, now also including the image solutes and thereby accounting for the reflected flow field of the surrounding solutes. In~\citet{Mereghetti2012}, $R_\textnormal{cut}= 4 a^{HI}$, i.e., four times the radius of the solute, was determined to be a good choice for crowded environments. Note that the flow field of a solute is reflected at the plane where the water molecules interact with the surface atoms, which is the atomic height of the surface (i.e., at z = radius of the surface atoms and not at z=0, where the centers of the surface atoms are positioned). The atomic surface height therefore has to be specified as an input parameter in SDA. 

\subsection{Grid-based Interactions}

In SDA, the forces between a pair of macromolecules $1$ and $2$ are calculated as finite-difference derivatives of the pairwise interaction, $\Delta G$, defined as:
\begin{align}
\begin{aligned}
\Delta G &= \frac{1}{2}\sum_{i_2}\Phi_{\textnormal{el}_1}(\R_{i_2})\cdot q_{i_2} + \frac{1}{2}\sum_{i_1}\Phi_{\textnormal{el}_2}(\R_{i_1})\cdot q_{i_1} \\
&+\sum_{i_2}\Phi_{\textnormal{ed}_1}(\R_{i_2})\cdot q_{i_2}^2\\
& + \sum_{i_1}\Phi_{\textnormal{ed}_2}(\R_{i_1})\cdot q_{i_1}^2 \\
&+\sum_{n_2}\Phi_{\textnormal{np}_1}(\R_{n_2})\cdot SASA_{n_2} \\
&+ \sum_{n_1}\Phi_{\textnormal{np}_2}(\R_{n_1})\cdot SASA_{n_1} \\
&+\sum_{n_2}\Phi_{\textnormal{rep}_1}(\R_{n_2}) + \sum_{n_1}\Phi_{\textnormal{rep}_2}(\R_{n_1})
\end{aligned}    
\label{eq:potential}
\end{align}
where $\R$ refers to the atomic coordinates of either atoms ($n_{1,2}$) or charges ($i_{1,2}$). A detailed description and parameterization can be found in references~\citenum{Gabdoulline2009}, \citenum{Mereghetti2010}, \citenum{Martinez2015} and \citenum{Gabdoulline1996}.
The first two terms in Eq.~\ref{eq:potential} represent the interaction energies of the charges  ($q_{i_2}$ or $q_{j_1}$) of one solute with the electrostatic potential ($\Phi_{\textnormal{el}_1}$ or $\Phi_{\textnormal{el}_2}$) of another solute~\cite{Gabdoulline1996}. 

To calculate the electrostatic interactions, the effective charge model~\cite{Gabdoulline1996} (ECM) is employed. These charges are calculated such that in a uniform dielectric they reproduce the  electrostatic potential in a shell around the solute that was previously computed by solving the Poisson-Boltzmann (PB) equation. Thereby, the effective charges implicitly account for inhomogeneities in the dielectric that do not have to be considered further during the BD simulation. Importantly, the required number of effective charges necessary to maintain a high accuracy is commonly much smaller than the number of partial atomic charges.

The third and fourth terms describe the electrostatic desolvation energy and account for the effects of the low dielectric cavity of one solute on the interaction energies of another~\cite{Gabdoulline1996}. These terms are computed as the interaction of the charges and the electrostatic desolvation potential of one another ($\Phi_{\textnormal{ed}_1}$ or $\Phi_{\textnormal{ed}_2}$)~\cite{Elcock1999} using the parameterization of reference~\citenum{Gabdoulline2009}. The fifth and sixth terms correspond to the non-polar interactions due to the burial of the solvent accessible surface areas (SASAs) of the surface atoms~\cite{Gabdoulline2009}. 

To avoid overlaps of solutes, two options are available in SDA: First, upon overlap, the BD step is repeated with a different random number until there is no overlap. For two-solute cases, such as the approach rate calculations for HEWL in section~\ref{sec:approach_rate}, this option provides a simple way to avoid overlap between solutes. For simulation systems consisting of many solutes, this option is not feasible, as overlaps occur much more frequently. Therefore, the second option, soft-core repulsive potentials with an inverse power function that hinders overlaps of solutes while at the same time preventing abrupt force changes at close contact are used in this case. These are described by the last two terms of Eq.~\ref{eq:potential}. 

In general, the potentials could be recomputed at every step of a simulation for the newly propagated configuration of atoms. In practice, we treat the macromolecules as rigid bodies (each having a single conformation or an ensemble of rigid conformations), allowing us to map all interaction potentials onto grids centered on each of the macromolecules that are only calculated once before the start of the BD simulation.

\subsection{Debye-H\"uckel Model for Long-range Solute-Solute Electrostatic Interactions}
\label{sec:dh}

For short-range interactions, such as van der Waals and desolvation forces, the interaction potentials decay quickly with distance, requiring only small grids to obtain a negligible truncation error. In addition, efficient storage algorithms, such as DT-Grid~\cite{Nielsen2006, Ozboyaci2016a}, exist that further lower the memory requirements for a given grid size. In contrast, for long-range electrostatic interactions,
algorithms such as DT-grid are not applicable
and using electrostatic potential grids sufficiently large to ensure the truncation error is small can result in both very high memory requirements and slow memory access times.

To alleviate the problem of the truncation of the electrostatic interaction grid at the boundaries, the long-range interactions can be captured by the Debye-H\"uckel (DH) approximation that was implemented in SDA by~\citet{Mereghetti2014} and~\citet{Martinez2015}. The method is well suited in combination with the effective charge approximation, since beyond the region where electrostatic interactions are treated on a grid-point charge basis, the energy term, $\Delta G_{DH}$, can simply be added to the energies given in Eq.~\ref{eq:potential}.  According to the DH theory of dilute electrolyte solutions, all ions in the solvent are treated as point charges while any pair of solutes $1$ and $2$ is treated as spheres with radii $a_1$ and $a_2$ and net formal charges $Q_1$ and $Q_2$. At a center-center separation of $r=|\R_1- \R_2|$, the additional DH energy term can be calculated as:
\begin{align}
\Delta G_\textnormal{DH}(r)= 
\begin{cases}
\infty,& \text{if } r < a^{DH}\\
\frac{Q_1 Q_2}{4\pi\epsilon_0 \epsilon_\textnormal{r}}\frac{\exp(-\kappa (r-a^{DH}))}{r(1+\kappa a^{DH})} & \text{if }   r \ge a^{DH}
\label{eq:two_solutes_DH}
\end{cases}
\end{align}
where ${a^{DH}=a^{DH}_1 + a^{DH}_2}$ is the sum of the radii of the two interaction solutes and ${\kappa = (\lambda_D)^{-1}}$ is defined as the inverse Debye length and is calculated as:
\begin{equation}
\kappa^2 = \frac{2e^2}{k_\textnormal{B} T \epsilon_0 \epsilon_\textnormal{r} }I
\end{equation}
where $\epsilon_0$ denotes the vacuum permittivity, $\epsilon_\textnormal{r}$ the relative permittivity of the solvent, $e$ the elementary charge, $k_\textnormal{B}$ the Boltzmann constant, $T$ the temperature, and $I$ the salt ionic strength. 

\begin{figure}[t!]
	\includegraphics[width=3.3in]{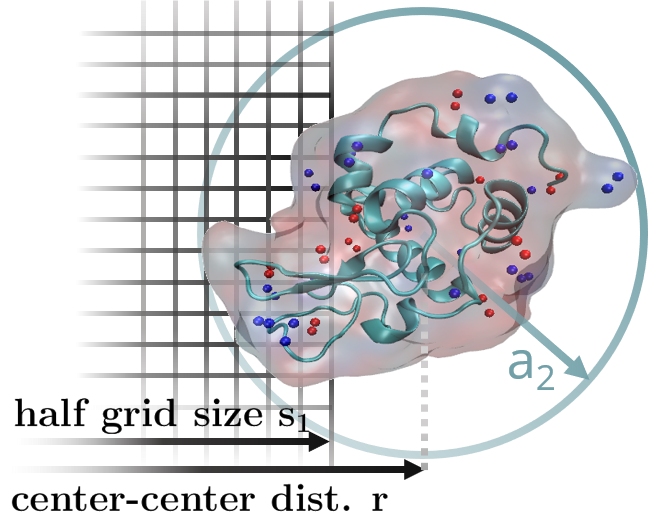}
	\caption{Illustration of the'partial' region of a solute electrostatically interacting with another solute or surface. By way of example, HEWL is shown in cartoon represntation with its molecular surface and with red and blue spheres representing the effective charges computed at pH~7. The effective charges in the 'partial' region that overlap with the potential grid of another solute or surface positioned to the left of HEWL are used to compute the grid-point charge interaction forces.} 
	\label{fig:transition_scaling}
\end{figure}

The full electrostatic potential grid is isotropically cut off to fulfill the assumption of centrosymmetry at the grid edges and to switch to the analytical DH potential beyond the cutoff distance. However, in the model described by \citeauthor{Mereghetti2014}, when a solute is on the border of the electrostatic potential of another solute, the grid-point charge interactions are calculated using only the fraction of its effective charges that is inside the grid, while the other charges are not considered. Furthermore, discontinuities in the forces may arise once the solute moves fully outside the potential grid, as the electrostatic calculation is switched to the DH treatment in a single step.
To improve the treatment of the transition between the full grid-based treatment and the region with only monopole interactions, a hybrid approach was introduced by Martinez et al.~\cite{Martinez2015}. This affected the region where only a fraction of the effective charges of the first solute is inside the electrostatic potential grid of the second solute, which we refer to  as the 'partial' region.  This region is illustrated for HEWL in Figure~\ref{fig:transition_scaling}.
The intersolute forces inside the partial region are computed from two contributions~\cite{Martinez2015}: For the effective charges located inside the grid, the partial interaction force $F_\textnormal{partial}(r)$ is calculated based on the effective charges interacting with the potential grid. Outside the grid, the DH treatment is applied but with the term computed for a solute charge equal to the sum of the effective charges lying outside the spherical grid boundary.

Here, we further modify the interaction model to improve the treatment of the partial region. This improved model is aimed at ensuring force and torque continuity, as well as consistency with the effective charge model. 
As before, for the effective charges located inside the grid, the partial interaction force $F_\textnormal{partial}(r)$ is calculated based on the effective charges interacting with the potential grid.
However, outside the grid, the use of a single charge (defined as a fraction of the effective rather than the formal net charge of the solute) for the DH calculation can lead to inconsistencies, e.g., in the calculation of the torque. Therefore, here, the effective charges are still employed for the calculation of the DH force $F_\textnormal{DH}(r)$, as they represent the geometry of the solute, but a scaling factor of total net formal charge / total effective charge is applied to each charge. As such, the scaled value of the effective charges leads to the correct formal net charge with the DH approximation accounting for the dielectric medium.

This procedure gives a smooth transition from the 'inside' region, in which one solute is fully inside the grid of another, to the 'partial' region, where it is partially outside the grid. The transition to the outside region (when the solute is fully outside the grid of the other solute) will, however, lead to a discontinuity in the forces and potentials, as the interaction is being reduced from a representation of many effective charges to a single charge in one step. Furthermore, on approach of two solutes, the abrupt appearance of effective charges may lead to artificial repulsive forces as, in the absence of intersolute torques, the solutes will not have been able to adjust to a favorable orientation with respect to each other. Therefore, we here introduce a linear scaling from the partial force $F_\textnormal{partial}(r)$ to the monopole interaction $F_\textnormal{DH}(r)$, within part of the partial region, 
\begin{align}
F(r) = \lambda_\textnormal{F}\, F_\textnormal{DH}(r) + (1-\lambda_\textnormal{F}) \,F_\textnormal{partial}(r) \;
\label{eq:scaling}
\end{align}
where the prefactor $\lambda_\textnormal{F}$ scales from zero at ${r=s_1-a^{DH}_2}$, to one at $r = s_1$, i.e. ${\lambda_\textnormal{F} = (r -s_1 + a^{DH}_2)\slash a^{DH}_2}$ for ${s_1 - a^{DH}_2 \leq r \leq s_1}$, where $s_1$ denotes the half-length of one side of the cubic electrostatic grid of solute 1. The split of the partial region into two parts ensures that the electrostatic force present upon a solute entering or leaving the grid of another is maintained while simultaneously avoiding discontinuities in the forces and numerical instabilities.

\subsection{Debye-H\"uckel Model for Long-Range Solute-Surface Electrostatic Interactions}
\label{sec:dh_surf}

Here, we describe our extension of the DH model for use with homogeneously charged surfaces. We assume an infinitely extended surface, which is locally well fulfilled under experimental conditions and a given in simulations if periodic boundary conditions are employed.  

Placing the surface without loss of generality at $z=0$, the corresponding potential at height $z$ is known from continuum electrostatics by solving the Debye-H\"uckel equation:
\begin{equation}
\frac{d^2}{dz^2}\Phi(z) = \kappa^2 \Phi(z)\;
\end{equation}
As the system is assumed to be uniform in the x and y directions, the potential only depends on the distance from the surface, $z$. Through an exponential ansatz under the boundary condition that the potential $\Phi(z) \rightarrow 0$ for $z \rightarrow \infty$, the solution
\begin{align}
\Phi(z) = \Phi_0 \: e^{-\kappa z}
\label{eq:ode}
\end{align}
is obtained, where $\Phi_0=\Phi(z=0)$. To determine $\Phi_0$, two different cases can be considered, depending on the system to be modelled. 

Firstly, we consider an infinitely thick surface with its charges located at the solute-solution interface and aqueous solution on one side and an infinite low dielectric region on the other. In this case, there is a surplus of counterions in solution to balance the net charge of the surface, i.e., the charge of the surface is equal and opposite to the charge of the solution. For this case, \citet{Grahame1947} calculated :
\begin{align}
\sigma = -\int_0^\infty \rho_\textnormal{e} \: \der z
\label{eq:charge_integration}
\end{align}
where $\sigma$ denotes the surface charge density and $\rho_\textnormal{e}$ is the volume charge density of the solution, with the condition of 
\begin{align}
\frac{\der \Phi}{\der z} (z\rightarrow \infty) = 0\;
\end{align}
He then used the differential form of Gauss' law to calculate $\rho_e$ and arrived at:
\begin{align}
\sigma &= \epsilon_\textnormal{r} \epsilon_0 \kappa \Phi_0 \: e^{-\kappa z} |_{z=0} \\
\Rightarrow \Phi_0 &= \frac{\sigma}{\epsilon_\textnormal{r} \epsilon_0 \kappa}
\label{eq:gouy_chapman}
\end{align}
This result is referred to as the Gouy-Chapman model. 

Now, we again consider the solute to be an extended sphere of radius $a$ with a low-dielectric region inside which the electrostatic potential does not decay exponentially.  Therefore, accounting for the sphere radius and substituting Eq.~\ref{eq:gouy_chapman} into Eq.~\ref{eq:ode}, we obtain: 
  
\begin{equation}
\Phi = -\frac{\sigma}{\epsilon_\textnormal{r} \epsilon_0 \kappa} \exp(-\kappa (z-a^{DH})) \;
\label{eq:gouy_chapman_extended}
\end{equation}

Secondly, we consider a thin surface with solution on both sides, i.e., with negligible thickness for long-range electrostatics. Here, it is assumed that the excess counterions balancing the charge of the surface in solution are equally distributed on both sides of the thin surface. Integrating Eq.~\ref{eq:charge_integration} from $-\infty$ to $\infty$, and again considering the solute as an extended sphere gives: 
\begin{equation}
\Phi = -\frac{\sigma}{2\epsilon_\textnormal{r} \epsilon_0 \kappa} \exp(-\kappa (z-a^{DH})) \;.
\end{equation}
Here, $\Phi_0$ differs from Eq.~\ref{eq:gouy_chapman_extended} by a factor of one half. Whether a thick or a thin surface is employed can be set in the input files of the SDA simulation package. 

\section{Simulation Details}

\subsection{Approach Rate Constant for Two Proteins}
\label{sec:approach_rate}

First, we assessed the effects of truncating the precomputed electrostatic potential grids used to model intersolute electrostatic interactions, and the performance of the intersolute DH model for correcting for this truncation. We simulated the diffusion of a pair of HEWL molecules and monitored the rate at which they first approach each other to within separations $d$ sampled at 1~\AA\ intervals in the range from 30 to 80~\AA. We compared the convergence of the approach rate constant computed from these three sets of simulations, with differing treatments of the long-range electrostatic interactions between solutes that lie partially or fully outside the potential grids of their interacting solutes, as the extent of the electrostatic potential grid was increased. Namely, we first used a simple truncation of the potential at the grid boundary with no DH correction at longer distances. Secondly, we used a truncation of the grids and the DH correction that considers the full effective charge distribution in the transition treatment at the grid boundary.  Thirdly, we used a truncation of the grids with the full transition treatment that includes the scaling of Eq.~\ref{eq:scaling}. 

For each grid extent and treatment of long-range electrostatics, 5,000,000 BD trajectories were performed in which a pair of HEWL molecules were initially separated by a center-to-center distance $b=600$~\AA, with both molecules rotated in different random orientations at the beginning of each trajectory. The trajectories were simulated until the proteins reached a separation of $c=900$~\AA. The fraction of trajectories $\beta(d)$ in which the two molecules diffused to a separation $d$ was monitored, allowing a rate constant for the diffusional approach of the two molecules to a separation $d$, which we name the approach rate constant $k_\textnormal{app}(d)$, to be calculated from: \cite{Northrup1983}
\begin{equation}
k_\textnormal{app}(d) = \frac{k(b)\beta(d)}{1-(1-\beta(d))\frac{k(b)}{k(c)}}
\label{eq:assoc_rate}
\end{equation}
where $k(b)$ and $k(c)$ are the rate constants for diffusion to the relative separations $b$ and $c$. These rate constants were calculated using the Smoluchowski approximation $k(x) \approx 4 \pi D x$, where $D$ is the sum of the diffusion coefficients of the two solutes. Note that the Smoluchowski equation requires the assumption that the force acting between two solutes at a distance $x$ can be assumed to be negligible. To check that the intersolute interaction energy is indeed negligible at the assigned $b$ and $c$ distances, we also calculated the results by determining $k(x)$ by numerical integration of
 \begin{equation}
 k(x) = \left(\int_x^{\infty}\mathrm{d}r\left[\frac{\exp(U(r)\slash k_\textnormal{B} T)}{4\pi r^2 D}   \right]\right)^{-1} \;,
 \label{eq:rate_const_full}
 \end{equation}
which only requires the assumption that the interaction force between the two solutes is centrosymmetric at separations of $b$ and larger. However, we found that the differences between the rates $k(b)$ and $k(c)$ calculated with Eq.~\ref{eq:rate_const_full} and with the Smoluchowski approximation were negligible.  

The crystal structure of HEWL\cite{Wilson1992} was taken from the RCSB Protein Data Bank (PDB code: 1HEL). The protonation states of ionizable amino acid residues were assigned with PDB2PQR\cite{Dolinsky2004} at pH 7.0 using the atomic partial charges and radii from the Amber force field, \cite{Cornell1996} giving a resulting net charge of +8 e on HEWL. Cubic electrostatic potential grids of length 96, 128, 160, 224, 352 and 480~\AA\ and grid spacing 1~\AA\ were created by solving the linearized Poisson-Boltzmann (PB) equation using the finite difference multigrid method implemented in APBS~1.4\cite{Baker2001} with an ionic strength of 5 mM, a solvent dielectric constant of 78.4, and a protein interior dielectric of 4.0. The dielectric interface was generated using a smoothed molecular surface and the potential at the boundary grid points was assigned using a multiple DH sphere model. Effective charges were calculated using the ECM module of SDA~7\cite{Gabdoulline1996,Martinez2015} by placing charge sites on the side chains of charged amino acid residues and the protein chain terminii. Potential grids describing electrostatic desolvation interactions were created using the make\_edhdlj\_grid module in SDA~7\cite{Martinez2015} following the parameterization described by Gabdoulline et al.\cite{Gabdoulline2009} 

The BD simulations were performed with a modified version of SDA 7 using a HEWL infinite dilution translational diffusion coefficient of 0.01232~\AA$^2\slash$ps and a rotational diffusion coefficient of $2.3057 \times 10^{-5}$~rad$^2 \slash$ps. Both diffusion coefficients were calculated with HYDROPRO \cite{Ortega2011}. During  the BD simulations, the intersolute repulsive interactions were modelled using an excluded volume grid of spacing 1~\AA\, generated for the HEWL structure using a probe radius of 1.77~\AA. Any attempted BD steps that resulted in an atom of the other HEWL molecule entering this volume were rejected and repeated with different random numbers. A HEWL radius of $a^{DH} = 15$~\AA, the radius of gyration calculated with HYDROPRO,\cite{Ortega2011} was used to describe the low dielectric cavity of the protein in the DH model. Due to the similarity of the radius obtained from X-ray scattering and from hydrodynamic measurements\cite{GarcadelaTorre2001, Ortega2011}, also $a^{HI} = 15$~\AA\ was used. 

\subsection{Adsorption of HEWL to a Mica Surface}

To assess the performance of the new solute-surface DH electrostatic model and the solute-surface HI models, we simulated the adsorption of HEWL to a mica surface at a range of HEWL concentrations in aqueous solutions of 5 and 50 mM ionic strength. The mica surface was approximated using a homogeneously charged graphite lattice surface with a charge density of $\sigma=-0.0215$ $e\slash$\AA$^2=-0.331$ C\slash m$^2$, corresponding to that of a mica surface at pH 7\cite{Shapiro2015}. This resulted in a charge of $ -0.053 $ \emph{e} on each graphite atom. The electrostatic potentials of this surface when surrounded by aqueous solvent of ionic strengths of 5 and 50 mM were calculated by solving the linearized PB equation with APBS~1.4.\cite{Baker2001} To approximate the low dielectric interior of a macroscopically-sized surface, 80 additional neutral graphite layers were stacked at separations of 1~\AA\ below the charged surface layer. All layers had planar dimensions of 502.68~\AA\ $\times$ 501.26~\AA. For each ionic strength, the linearized PB equation was solved using a coarse cubic grid of length 1056~\AA\ with 353 grid points in each dimension. The other calculation parameters were as described for the HEWL calculation above. Following this, two finely-spaced electrostatic potential grids of different sizes were computed using the coarse grid to provide a focusing boundary condition. Both grids had spacings of 1~\AA\ and lengths of 352~\AA\ in the dimensions in the plane of the surface.  The larger grid had a total length of 352~\AA\ in the dimension perpendicular to the surface and extended 336~\AA\ above the surface. The smaller grid had a total length of 128~\AA\ and extended 114~\AA\ above the surface. For simulations at 50 mM, additional HEWL electrostatic potential grids and effective charges corresponding to this ionic strength were created following the procedure described above. In all the simulations performed in the presence of a surface, we used the HEWL electrostatic potential grid with sides of length 96~\AA. We show below that, when combined with the intersolute DH correction, this grid size is sufficient to accurately represent HEWL electrostatic interactions. 

In SDA, the electrostatic interactions between a pair of interacting species are approximated using the effective charge model\cite{Gabdoulline1996} as the mean of the interactions due to the set of effective charges on each interacting partner with the PB-derived electrostatic potential on the other (the first two terms in Eq.~\ref{eq:potential}). These two terms should be approximately equal. To account for the desolvation of solute-facing charges on each interacting partner as the two species approach each other closely, additional correction terms are applied\cite{Elcock1999} (the third and fourth terms in Eq.~\ref{eq:potential}). This effective charge model is able to approximate the PB-calculated electrostatic interaction forces between the partners at each time step of the simulation, without requiring repeated solution of the PB equation. However, it is more difficult to define appropriate effective charges on an infinite surface by fitting the electrostatic potential within a given shell around the surface. For a planar surface, Romanowska et al.\cite{Romanowska2015} found that the effective charges calculated on the surface had to be scaled so that the first two terms in Eq.~\ref{eq:potential} were similar in magnitude for docked HEWL configurations in which the protein was close to the surface. Here, to describe the adsorption process, we need to compute the electrostatic interactions between HEWL molecules and the surface to good accuracy at all heights above the surface. As the scaling used in Romanowska et al. is not able to do this, we instead calculated the surface -- HEWL electrostatic interaction in one direction only, i.e. using the electrostatic potential grids on the surface and the effective charges on the HEWL molecules, effectively replacing the first two terms of Eq.~\ref{eq:potential} with a single term with no $\frac{1}{2}$ prefactor. Therefore, effective charges were not computed for the mica or silica surfaces modelled in this work.

The interaction grids for defining the electrostatic desolvation potential of the surface were calculated with make\_edhdlj\_grid from SDA~7\cite{Martinez2015} following the parameterization described by Gabdoulline et al.\cite{Gabdoulline2009}. The electrostatic desolvation grid of HEWL described in the previous section was used.  The additional grids describing the non-polar desolvation potentials of the surface and HEWL were calculated following the parameterization used in the graphite -- hydrophobin simulations described by Mereghetti and Wade.\cite{Mereghetti2011}  Unlike the two solute simulations described in the previous section, repulsive interactions were modelled using a soft-core repulsive model, again following the parameterization used for the graphite -- hydrophobin simulations\cite{Mereghetti2011}. The parameterization of all interaction terms used in these simulations were chosen to be consistent with those used in previous simulations of HEWL solutions.\cite{Mereghetti2014}

\afterpage{
	\bgroup
	\def\arraystretch{1.3}
	\begin{table}[t!] 
		\centering
		\begin{tabular}{c|c}
			HEWL Protein & \\
			Concentration [g\slash l] & Box height [\AA] \\
			\hline
			5  &  6837  \\
			10  & 3418 \\
			20  & 1708 \\
			30  &  1138
		\end{tabular}
		\caption{Heights of the simulation boxes chosen for the HEWL concentrations used to study the adsorption of HEWL to a mica surface. For all heights, the quadratic base area of the periodic simulation box was (322.8~\AA)$^2$.}
		\label{tab:adsorption_box_sizes}
	\end{table}
	\begin{figure}[h!] 
		\centering
		\includegraphics[width=1.0\linewidth, angle=0]{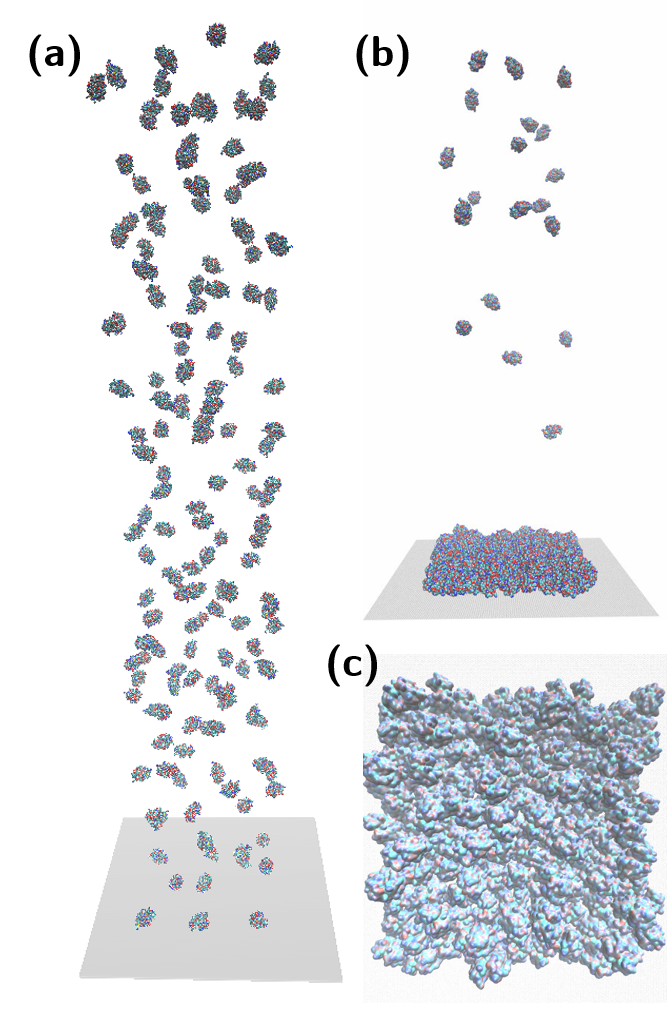}
		\caption{Snapshots of a simulated system consisting of 150 HEWL proteins in a periodic simulation box at a concentration of 30 g\slash l adsorbing to an oppositely charged mica surface. (a) Initial starting configuration, (b) side view and (c) top view after 12~$\mu$s simulation time. Images created with VMD \cite{Humphrey1996}. }
		\label{fig:sim_setup_hewl}
		\vspace{2em}
	\end{figure}
} 

BD simulations, each of a duration of 12~$\mu$s, were performed with a time step of 0.4 ps at HEWL concentrations of 5, 10, 20 and 30 mg/ml. Note that, as the experiments were performed on timescales from minutes to hours and at concentrations in the range of ${\mu\mathrm{g}\slash\mathrm{ml}}$\cite{Daly2003a, Robeson1996}, it is not feasible to simulate these systems in atomic detail at the experimental time and length scales. Therefore, we used higher concentrations than in the experiments, that were chosen with the aim of reproducing the qualitative trends observed in experiments and providing insights into the mechanisms of the adsorption processes and the contributions of the different components of the interaction forces.

For each HEWL concentration, an initial configuration of 150 HEWL molecules was placed in a box with periodic boundaries and length 322.8~\AA\ in both dimensions parallel to the surface. These dimensions were chosen to be significantly smaller than the surface used for generating the PB electrostatic grid in order to minimize the curvature of the isoelectric field contours, so that the electrostatic potential calculated in the non-periodic PB calculation provided a good approximation of that of an infinite charged plane in the simulated volume. The extents of the simulated volumes above the surface were chosen to give the desired HEWL concentrations (Table~\ref{tab:adsorption_box_sizes}). For a HEWL concentration of 30~mg$\slash$ml, Figure~\ref{fig:sim_setup_hewl}a shows a snapshot of the starting configuration. The region within 120~\AA, i.e., four times the diameter of HEWL, above the surface was left empty to avoid an immediate collapse of a number of proteins onto the surface that would not occur at experimental conditions at lower concentrations.  Figure~\ref{fig:sim_setup_hewl}b and c show snapshots from different perspectives during the simulation process (again for 30~mg$\slash$ml HEWL concentration) after the majority of proteins adsorbed to the surface. 

A protein was considered to be adsorbed if its geometric center was located less than 50~\AA\ above the surface. This criterion was chosen because, considering the size and the ellipsoidal shape of HEWL, it captures two adsorption layers (see Figure~\ref{fig:density_map} and corresponding discussion). We tested that the results were robust with respect to the choice of this criterion, and that the proteins positioned below this threshold were stably associated with the surface and not diffusing freely. 

At each HEWL concentration, four simulations were performed for a solution of 5 mM  salt concentration. The first three, in which HI were ignored, varied in the treatment of long-ranged electrostatic interactions: (1) the larger electrostatic potential grid, which extended 336~\AA\ above the surface, with the charged surface DH correction above this and intersolute DH correction for solute -- solute interactions that extend beyond grid boundaries; (2) the smaller electrostatic potential grid that extended 114~\AA\ above the surface and both DH corrections; and (3) the smaller potential grid with no corrections. In the fourth simulation, HI were accounted for and the smaller electrostatic potential grid was used with both DH corrections.  

Two further simulations were performed at each concentration assuming a solution of 50 mM  salt concentration, one neglecting and one accounting for HI. In both of these simulations, the smaller electrostatic potential grids and both DH corrections were used. As we wish to model a macroscopically-sized charged surface, we represented it as an infinitely thick solvent-excluding surface, i.e. using Eq.~\ref{eq:gouy_chapman}. The HEWL radius used for calculating HI was 15~\AA\, which we also used for the solvent excluded region in the DH models. All other simulation parameters were as described for the two-HEWL simulations.

\sloppy
Lastly, as a comparison, another set of simulations with a silica surface was conducted at the HEWL concentrations mentioned above. Again, silica was approximated using a homogeneously charged graphite lattice surface, but with a charge density of
${\sigma=-0.0013}$~\mbox{e$\slash$ \AA$^2$}=${-0.02~\mathrm{C}\slash \mathrm{m}^2}$, corresponding to that of a silica surface at pH~7\cite{Daly2003a}. This assignment resulted in a charge of $-$0.0032~$\mbox{e}$ on each graphite atom. Here, the simulations were only conducted at 5~mM salt concentration using an electrostatic potential grid that extended 114~\AA\ above the surface and was solved as described above for the mica surface. The DH correction and HI were included. All other interactions were calculated as for the mica surface.

\section{Results}

\subsection{Convergence of the approach rate constant for two HEWL proteins with increasing electrostatic potential grid size}

The rate constants for the approach  of two HEWL proteins to separations in the range from 30 to 80 \AA\ were calculated using three different models for electrostatic interactions, and the convergence of these models with increasing electrostatic potential grid size was determined (Figure~ \ref{fig:dh_hewl_self_assoc_rate}). Using the first model, a simple truncation of the electrostatic interactions when the effective effective charges of one protein lie outside the electrostatic potential grid of the other, large truncation errors were seen when using the smaller grids (Figure~ \ref{fig:dh_hewl_self_assoc_rate}a). Only when using the second largest grid, which included the potential within a radius 176~\AA\ from the protein center, did the rate constant curve converge to that of the larger grid with a 240~\AA\ radius. When using smaller grids, considerably higher rate constants were computed due to the lack of electrostatic repulsion between the HEWL proteins, which have identical net charges, at longer distances.

\begin{figure}[t!]
	\includegraphics{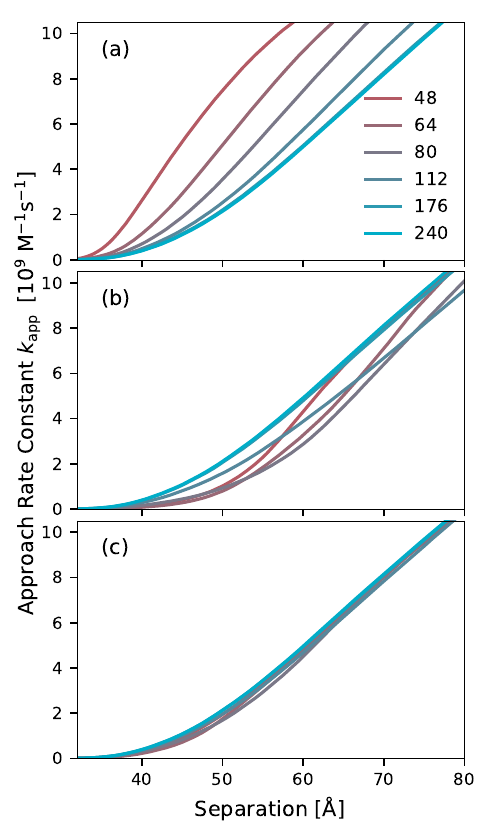}
	\caption{Approach rate constant for two HEWL protein solutes. Simulations were performed with six sizes of the electrostatic potential grids centered on each of the two proteins as indicated by the colors and the corresponding radii in~\AA. (a) No long-range electrostatic interactions beyond the grid boundaries were considered. (b) Using the DH approximation beyond the grid radius with the transition treatment as in \citet{Martinez2015}, but modified so that the charge  $Q$ for each solute in Eq. \ref{eq:two_solutes_DH} is defined by the formal charge of the protein, rather than the sum of its effective charges. (c) Using the full improved transition treatment with a smooth scaling switch (Eq.~(\ref{eq:scaling})) between the grid-based and Debye-H\"{u}ckel regimes.}
	\label{fig:dh_hewl_self_assoc_rate}
\end{figure}

While the simulations performed with a simple truncation of the electrostatic interactions showed a clear convergence of the approach rate constant for all separations as the grid size increased (Figure \ref{fig:dh_hewl_self_assoc_rate}a), this was not the case when the transition treatment described in \citet{Martinez2015}, with the corrected assignment of solute net charges $Q$ but not the scaling of Eq.~(\ref{eq:scaling}), was used, although  the errors in these simulations were generally of smaller magnitude than obtained with simple truncation (Figure~ \ref{fig:dh_hewl_self_assoc_rate}b). The errors in these simulations had the opposite effect to those obtained with the truncation model, leading to a decrease in the approach rate constant at all separations. We again observed that only the rate constants from simulations using the second largest grid  (with a 176 \AA\ radius) agreed with those obtained using the largest grid at all separations. The smallest grid produced the largest error in the predicted rate constants at smaller separations (Figure \ref{fig:dh_hewl_self_assoc_rate}b), whereas the third largest grid (with a 112 \AA radius) produced the largest error in the predicted rate constants at larger separations. The reason for these errors is apparent when examining the rate constants predicted for the simulations performed using the two smallest grid sizes. At larger separations, the rate constants match well with those of the simulations performed with the 240 \AA\ radius grids, showing that, when corrected to use the formal charge of the proteins in place of the sum of their effective charges, the Debye-H\"{u}ckel model accurately models electrostatic interactions at these longer distances. It is only when the effective charges on the two HEWL proteins begin to enter into the electrostatic potential grids of the other protein, at separations slightly larger than the grid radius, that the predicted rate constants diverged. The abrupt entrance of the effective charges into the electrostatic grid of the interacting protein means that the proteins did not have the opportunity to rearrange into more favorable orientations, leading to unphysical repulsions in this region, and resulting in a reduction in the calculated rate constants.

The simulations run with the smoothed transition between the grid-based and Debye-H\"{u}ckel regimes showed much reduced errors for all grid sizes at all separations (Figure~\ref{fig:dh_hewl_self_assoc_rate}c).  Even with the smallest 48 \AA\ radius grids, the divergence in the predicted rate constants from those obtained from the simulations with the largest grid size was minimal. For this reason, the 48 \AA\ radius grid was used in all subsequent simulations, as it can accurately describe the electrostatic interactions of HEWL when coupled with the new Debye-H\"{u}ckel model.

\subsection{Adsorption of Multiple HEWL Proteins to a Mica Surface}
 
 \begin{figure*}[t!]
 	\includegraphics[width=\linewidth]{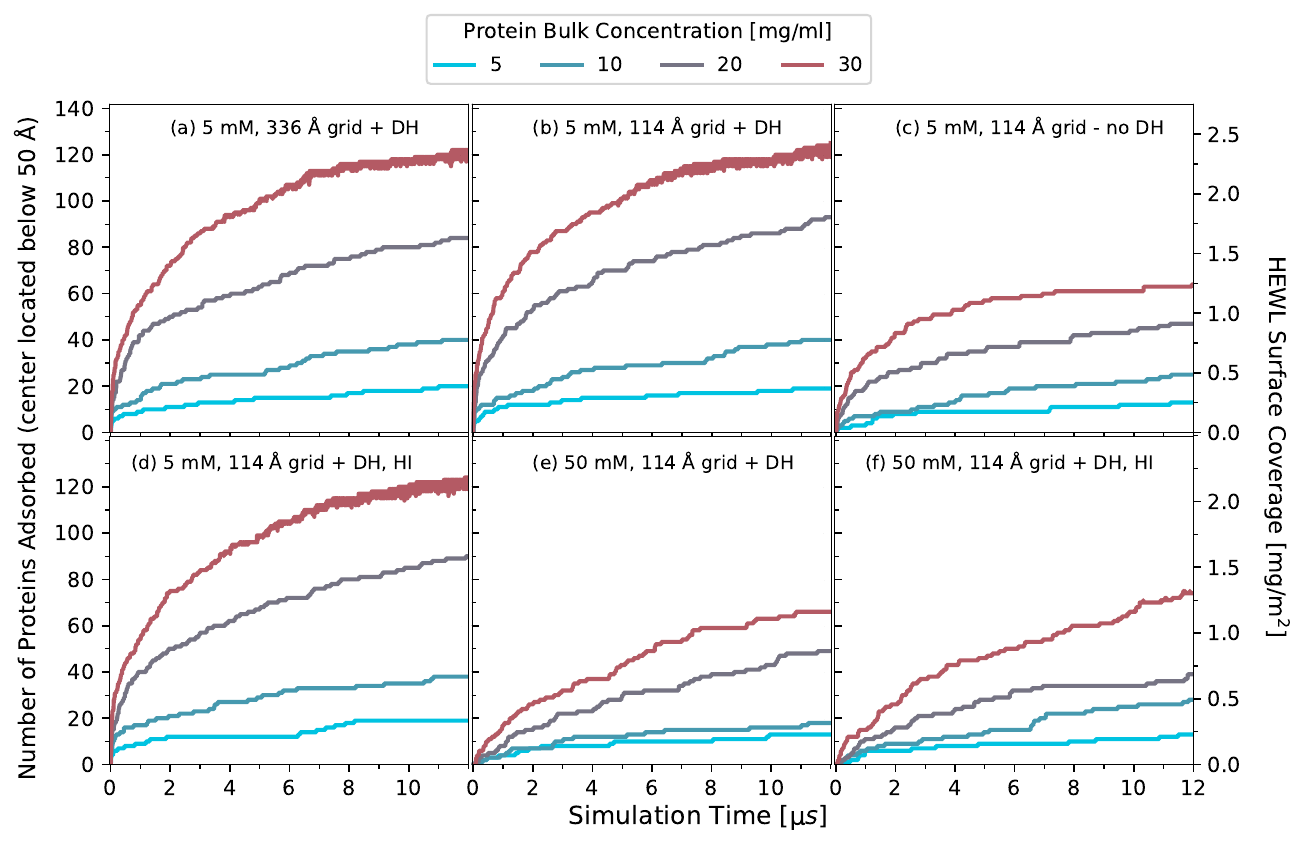}
 	\caption{Simulation of HEWL adsorption to a mica surface. Protein adsorption in simulations with different initial bulk protein concentrations (5 - 30~mg$\slash$ml) are shown by color and are compared for two different salt concentrations (5 (a-d) and 50 (e-f) ~mM) and different treatments of electrostatic interactions and HI. (a) - (c) show the results from different treatments of electrostatic interactions, comparing the use of a large electrostatic potential grid (336~\AA\ height above the surface)(a) to a smaller one (114~\AA) with (b) and without an additional long-range DH treatment (c). The effect of including solute-solute and solute-surface HI is shown in  (d) and (f).}
 	\label{fig:adsorption_HEWL_comb}
 \end{figure*}
 
 In all the simulations of systems consisting of multiple HEWL proteins and a mica surface, which were performed with several simulation conditions and interaction treatments, we observed that both the number of adsorbed proteins and the corresponding surface coverage in ${\mathrm{mg}\slash\mathrm{m^2}}$ increased with simulation time, with the adsorption rate being faster initially and then gradually levelling off, see Figure~\ref{fig:adsorption_HEWL_comb}. However, the number of adsorbed proteins was highly dependent on the initial bulk protein concentration. 

The simulations at 5~mM salt concentration with the electrostatic interactions calculated using a grid with a height of 114~\AA\ above the surface and the interactions beyond the grid captured with the DH model (Figure~\ref{fig:adsorption_HEWL_comb}b) yielded similar surface coverage values at all HEWL concentrations to those obtained when using a larger electrostatic potential grid extending 336~\AA\ above the surface together with the DH approximation beyond the grid (Figure~\ref{fig:adsorption_HEWL_comb}a). When the DH treatment to correct for the truncation error was not used, the resulting surface coverage was greatly reduced (Figure~\ref{fig:adsorption_HEWL_comb}c). 

At the higher salt concentration of 50~mM (Figure~\ref{fig:adsorption_HEWL_comb}e) and otherwise the same simulation conditions as in Figure~\ref{fig:adsorption_HEWL_comb}b, the results remain similar for low HEWL concentrations but differ for higher concentrations. The number of adsorbed proteins after 12~$\mathrm{\mu}$s of the simulations at 30~mg$\slash$ml protein concentration is almost half that at 5~mM. In contrast, at the lowest protein concentration of 5~mg$\slash$ml, the number of proteins adsorbed at the two salt concentrations is very similar.

\begin{figure}[t!]
	\includegraphics[width=3.33in]{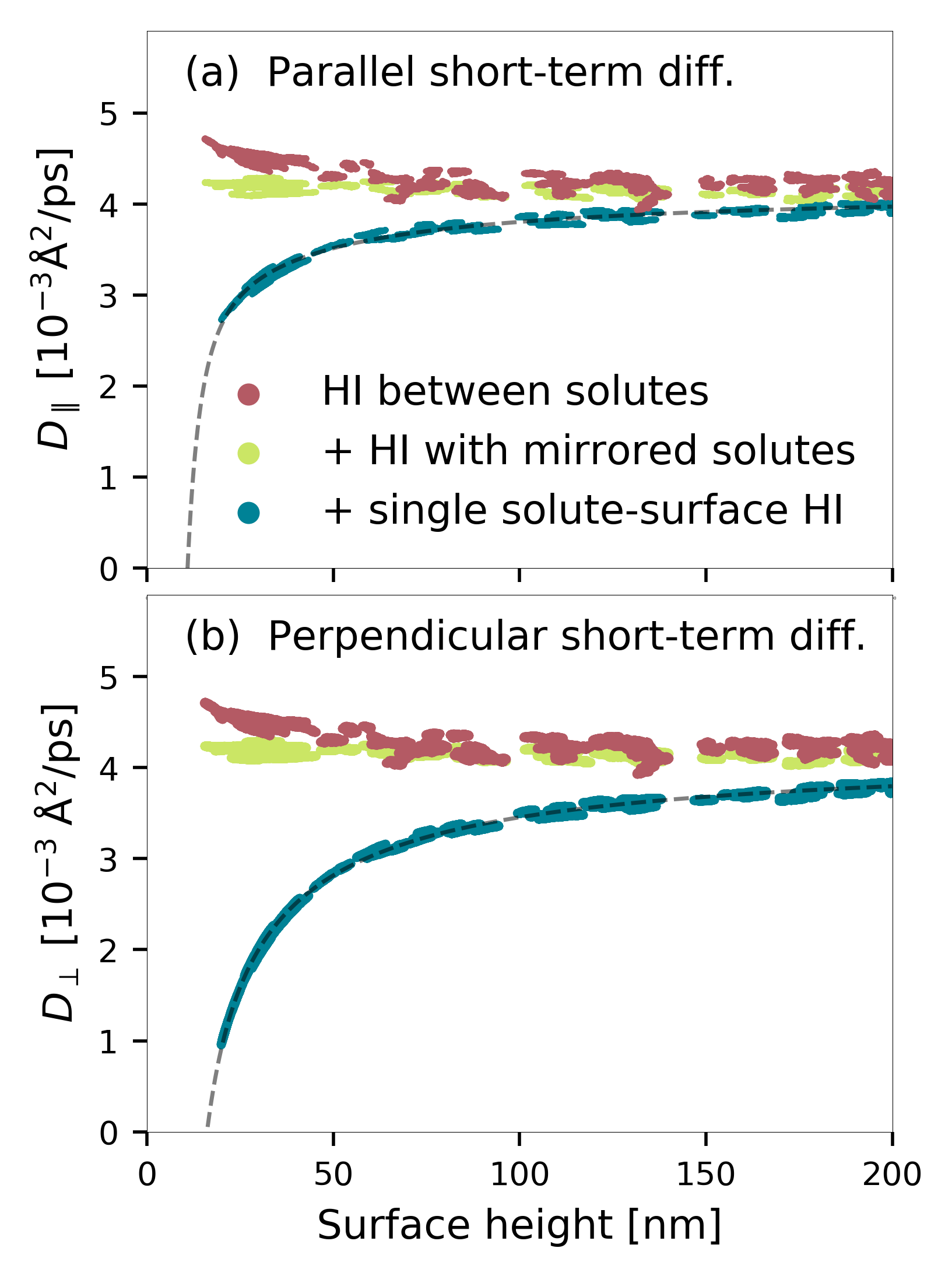}
	\caption{Short-time diffusion coefficients of the HEWL proteins computed from BD simulations with different HI contributions included. Each dot represents a protein. For the simulation conditions of Figure~\ref{fig:adsorption_HEWL_comb}d, the short-time coefficients of the individual proteins are plotted against the height of their center above the surface. The red dots indicate the results when  HI between the proteins only were considered, green when, in addition, the image flow field of the surrounding proteins was considered, and blue when also accounting for the HI of the protein with its own reflected flow field. The dashed line shows the analytical model of the single solute-surface HI after taking HI between solutes and mirrored solutes into account (i.e., multiplying the average diffusion coefficient indicated by the green points with the analytical reduction coefficient)}
	\label{fig:hi_contributions}
\end{figure}

Considering HI between the proteins, as well as between the proteins and the surface (Figure~\ref{fig:adsorption_HEWL_comb}d and f, at 5~mM and 50~mM salt concentration, respectively), the results are similar to those from corresponding simulations with HI neglected (Figure~\ref{fig:adsorption_HEWL_comb}b and e, respectively). Furthermore, the simulations shown in Figure~\ref{fig:adsorption_HEWL_comb}d were extended to 30~$\mu$s simulation time (see section~2 of the Supporting Information). Whereas the rate of protein adsorption slows and the number of adsorbed proteins levels off, the results show that the adsorption process continues beyond this time scale.

\subsubsection{Adsorption Patterns}
Examination of the distribution of the proteins revealed that two protein layers can form on the mica surface during the simulations. This distribution is shown in Figure~\ref{fig:density_map} at the end of the simulation with a 30~mg$\slash$ml protein concentration and the simulation conditions of Figure~\ref{fig:adsorption_HEWL_comb}d. 

We next evaluated the short-time diffusion coefficients of the proteins during the assembly of the first adsorption layer. For the simulations at a protein concentration of 30~mg$\slash$ml (Figure~\ref{fig:adsorption_HEWL_comb}d), scatter plots of the diffusion coefficients parallel and perpendicular to the surface are shown in Figure~\ref{fig:hi_contributions}a and b, respectively, against the height of the protein's center above the surface, with one point representing one solute. When using the mean-field approach to account for the HI between the solutes only, the short-time diffusion coefficient increases as the proteins approach close to the surface (red dots). The reason is that this method considers the part of the space occupied by the surface as empty. Extending the mean-field approach by taking into account image solutes beyond the surface, also accounts for the flow field reflection of the surrounding solutes, and resulted in the diffusion coefficients of the proteins being roughly constant at all distances from the surface, as shown by the green dots. However, when the HI computation included the direct coupling effects of the solutes with their own reflected flow field, there was a considerable decrease of the short-time diffusion coefficents close to the surface, as shown by the blue dots.

\begin{figure}[t!]
	\includegraphics[width=3.33in]{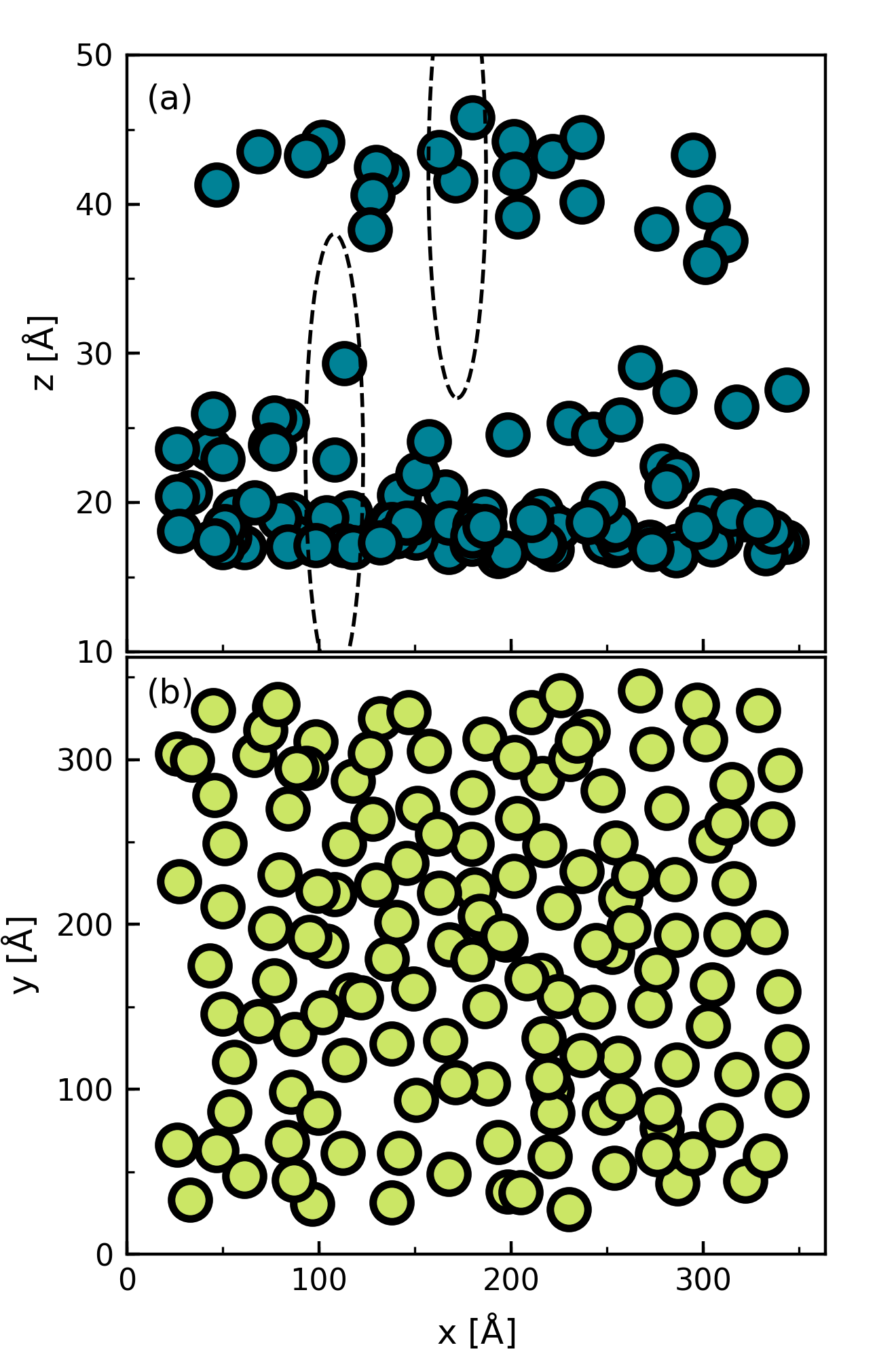}
	\caption{Adsorption patterns of HEWL on the mica surface at the end of a BD simulation (i.e. after 12~$\rm{\mu}s$),  at 30~mg$\slash$ml HEWL concentration. The circles indicate the positions of the centers of geometry of the proteins and are viewed (a) from the side and (b) from the top. The circles are smaller than the size of HEWL to faciliate visualization. In (a), the two dashed lines indicate the projection of a sphere with the radius of gyration of HEWL. They are distorted, as the distribution of the proteins in the $x$ and $z$ directions is shown on different scales. }
	\label{fig:density_map}
\end{figure}

\subsection{Adsorption to a Silica Surface}

We next conducted simulations with a silica surface at 5~mM salt concentration using an electrostatic potential grid extending 114~\AA\ above the surface and the DH approximation beyond. The full HI model was used for the simulations. In comparison to mica, the number of adsorbed proteins at all four HEWL concentrations simulated was lower for the silica surface (Figure~\ref{fig:silica}), due to the lower surface charge of silica compared to mica. Furthermore, in comparison to mica, the adsorption curves levelled off much earlier, especially at the higher protein concentrations. The number of proteins in the plateau region of the simulations at 20 and 30~mg$\slash$ml bulk protein concentrations is more similar, although it was slightly higher at the higher concentration.  

\begin{figure}[t!]
	\includegraphics[width=3.33in]{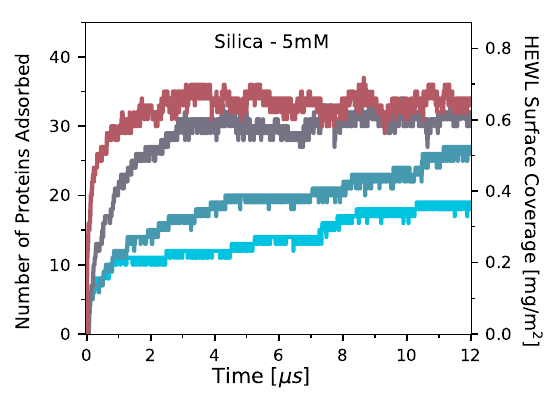}
	\caption{Simulation of HEWL adsorption to a silica surface at four bulk concentrations of HEWL (5, 10, 20 and 30 mg/ml). Apart from replacing the mica surface with a silica surface, the simulation conditions and the treatment of interactions, as well as the color scheme, are the same as for Figure~\ref{fig:adsorption_HEWL_comb}d.}
	\label{fig:silica}
\end{figure}

\section{Discussion}

We first consider our results with respect to the newly introduced methods, and then consider the insights into the systems studied that can be obtained from the simulations. 

We computed the approach rate constants characterizing the rate at which two HEWL molecules reach a certain distance from one another by diffusion. As described in section~\ref{sec:dh}, the treatment of long-range electrostatic interactions between two solutes has been improved with respect to the transition treatment between the grid based full charge representation and a one-dimensional DH approximation. These improvements lead to accurate results for three dimensional interaction grids that are much smaller --- by almost a factor five in the case of HEWL --- than previously required for the same accuracy. 

The reproduction of the adsorption of HEWL proteins to a mica surface required the simulation of a high number of molecules --- here, 150 HEWL proteins --- over several microseconds to cover the full adsorption process. The DH approximation for homogeneously charged surfaces, together with the improved treatment of the solute electrostatic grid transition, was used to capture the long-range electrostatic interactions. The agreement between the results obtained by using grids of 336 and 114~\AA\ height (Figure~\ref{fig:adsorption_HEWL_comb}a and ~\ref{fig:adsorption_HEWL_comb}b, respectively), indicate that also in this case, a full charge representation is only necessary at close distances when using the improved DH approximation for long-range interactions. Furthermore, even though the electrostatic interaction decays exponentially, 
the much smaller numbers of adsorbed proteins observed when the interactions beyond the grid are neglected (Figure~\ref{fig:adsorption_HEWL_comb}c) shows that the long-range electrostatic forces still critically influence the long-time diffusion behavior.

From a computational perspective, the reduction in memory and storage requirements achieved by using the improved long-range DH tratement becomes important when simulating the diffusion of large macromolecules or when carrying out simulations with large numbers of different solute molecules, each of which has its own precomputed electrostatic potential grid. 

We furthermore extended the computationally efficient mean-field treatment of HI\cite{Mereghetti2012} for many molecules to include confining surfaces. Both the relative anisotropic reduction in diffusion due to solute-surface HI, as well as the reduction based on the HI with the reflected flow-field of the surrounding solutes are included, in addition to the direct effect of HI between solutes. It is important to note that, due to the complexity of HI, these effects are combined in a first-order approximation, thereby neglecting higher order terms. 

For HEWL adsorption, although HI largely decreased the mobility of the solutes in the vicinity of the surface (Figure~\ref{fig:hi_contributions}), the HI showed little to no effect on the adsorption curves at 5~mM (Figure~\ref{fig:adsorption_HEWL_comb}d) and 50~mM (Figure~\ref{fig:adsorption_HEWL_comb}f) salt concentration. Similarly to other studies, the results show that while HI predictably change the local properties of a system, it is generally hard to predict the effect of HI on averaged global observables \textit{a priori}. \citet{Antosiewicz2017} analyzed the kinetics of diffusional encounters of, among others, Barnase and Barstar, and found that, even though the magnitudes of the torques resulting from the hydrodynamic coupling of the associating molecules were comparable with the magnitudes of the torques from electrostatic interactions, the overall effects of the hydrodynamic torques on the association kinetics were rather small. For studies of solutes in the presence of a surface, there is agreement that the diffusivity of the solutes is reduced, although there is a strong dependence on the properties and assumptions of the individual systems studied as regards whether this reduction also influences the adsorption properties. For irreversible adsorption processes of spheres to an attractive surface, \citet{Pagonabarraga1994} found that although HI influences local properties such as the pair correlation,  the macroscopic quantities were largely unaffected. In contrast, studies on similar models revealed that the time dependence of the surface coverage near saturation\cite{Wojtaszczyk1998} and the saturation level\cite{Perez2015} were highly influenced by the HI between the free solutes and the adsorbing surface. 

Importantly, these models often only consider adsorption as a purely diffusive process, e.g., as in the studies based on the Random Sequential Adsorption models \cite{Schaaf1989}. Therefore, changes in diffusivity influence both the approach to the surface and the task of finding a free spot on the surface. While some models include an attractive force to the surface \cite{Pagonabarraga1994, Wojtaszczyk1998}, they generally do not consider interactions between the solutes beyond exclusion and HI. As these neglected interactions are often strong for protein interactions, the results for proteins can deviate highly from the predictions of these models.

The simulation results can be compared to a number of experimental studies \cite{Daly2003a, Robeson1996, Su1998a, Lu2004}. However, it should be recalled that, firstly, we simulate bulk concentrations of several ${\mathrm{mg}\slash\mathrm{ml}}$ compared to ${\mathrm{\mu g}\slash\mathrm{ml}}$ in the experimental studies. Secondly, the results often vary between different experimental techniques\cite{Su1998a, Lundin2010}, and even between different setups for the same technique \cite{Kim2002, Lu2004, Lundin2010}. Therefore, the simulations  described here are only intended to reproduce qualitative trends, but can thereby provide insights into the adsorption processes. 

We observe that the simulated adsorption kinetics strongly depend on the protein concentration, which is in agreement with experimental studies \cite{Daly2003a, Robeson1996, Su1998a, Lu2004}. Furthermore, the adsorption kinetics are dependent on the salt concentration, as, due to the change in the decay of the electrostatic potentials, the repulsive forces between already adsorbed proteins and newly approaching ones becomes more dominant in comparison to the attractive forces with the surface. This observation is again in agreement with experiments\cite{Daly2003a}. 

The conditions of the adsorption simulation to a mica surface match experimental studies conducted at 5~mM ionic strength by \citet{Daly2003a} using, among other techniques, streaming current measurements to investigate the saturated surface coverage, based on the adsorption mechanisms proposed by \citet{Robeson1996} under the same conditions.
The surface coverage obtained of 2.3~${\mathrm{mg}\slash\mathrm{m^2}}$ is slightly higher than in the experimental studies (2.0~${\mathrm{mg}\slash\mathrm{m^2}}$). However, the reported amount was observed in the first layer on the surface. If we only consider the first layer (adsorption curves not shown), we obtain around the same value of 2.0~${\mathrm{mg}\slash\mathrm{m^2}}$. 
Especially for the mica surface, however, it is not possible to identify whether the adsorption curves for different concentrations will converge to the same level, as the levelling off is very slow at the lower concentrations. Even for the simulations extended to 30~$\mu$s (Supporting Information, section~2) a plateau region was not reached, particularly at the lower concentrations. Simulating even higher concentrations would, however, further increase the artificial effect of repulsion between the proteins in the bulk that pushes them towards the surface, which does not occur to such a degree in experiments at more dilute conditions. 

The qualitative features of the adsorption kinetics to the silica surface match those for the mica surface. The adsorption curves for the two highest concentrations appear converged at the end of the simulations. However, it is unclear if the small remaining difference in the number of adsorbed proteins between the two plateaus is due to additional repulsion between the proteins in bulk, or if it would vanish after a much longer simulation time. Two experimental studies using neutron reflection\cite{Su1998a} and dual polarization interferometry\cite{Lu2004} reported surface coverages of 3.29, 2.10 and 0.74 ${\mathrm{mg}\slash\mathrm{m^2}}$ for protein concentrations of 4, 1 and 0.03 ${\mathrm{mg}\slash\mathrm{ml}}$, respectively, so our result is slightly higher than that for the lower protein concentration. As the final surface coverage is much lower than that obtained by just considering the first layer at the mica surface, this shows that the adsorption to the silica surface is not limited by available space on the surface, but rather regulated by the electrostatic interactions between the proteins and the surface. In contrast, for the higher protein concentrations, the neutron reflection study reports several adsorption layers, which we, however, only observe for the more highly charged mica surface. 

Two additional approximations underlying the BD simulations presented here should be emphasized again. Firstly, HEWL is treated as a rigid body. While HEWL is not known to show large conformational transitions upon adsorbing to a surface\cite{Daly2007}, flexible tail and side-chain movements may facilitate binding to the surface\cite{Steudle2011}, and hence, may result in a denser packing of the proteins on the surface. For a small number of flexible proteins, the BD simulations could be combined with a  MD simulation technique to account for induced fit in the last step of the adsorption, as shown in Ref.~\citenum{Ozboyaci2016}. Secondly, in the BD simulations of systems of many solutes and a surface, the forces acting on one solute are obtained by summing the contributions from the interactions the surrounding solutes and the surface in an additive manner. It is thereby assumed that the linearized PB equation holds, which may not be fully justified for highly charged species, e.g., for the mica surface without any oppositely charged proteins adsorbed to it.  
With these assumptions, the combined methodological advances described here will enable the computationally efficient study on other adsorption processes or of systems with confined crowded environments that are hard to access \textit{in vivo} experimentally. A BD simulation of 12~$\mu$s with over 150 HEWL molecules treated in atomic detail for one bulk protein concentration took less than a week on a compute node with 16 CPU cores. 

\section{Conclusions}

We have here presented new computational models to efficiently capture hydrodynamic and long-range electrostatic interactions for BD simulations of proteins in the presence of a surface. These models have been implemented in the SDA software package.  Proteins and surfaces are modelled at full atomic detail while approximating them by one or more rigid bodies. Considering the self-association of HEWL and the adsorption process of many HEWL molecules to mica and silica surfaces, long-range electrostatic interactions were shown to critically influence the diffusional kinetics at low ionic strength, but, with the improved methodology described here, only had to be calculated in full detail at close distances. In contrast, while HI lowered the mobility of the proteins close to the surface, they had almost no effect on the observed protein adsorption kinetics. The simulations were able to reproduce trends in protein-surface adsorption properties observed in different experimental conditions. In conclusion, the methodology presented here enables the simulation of over a hundred proteins in the presence of a surface modelled in atomic detail at very low computational cost compared to, for example, atomistic explicit solvent MD simulations.

\section{Code and Data Availability}
The SDA software package is available at \url{http://mcm.h-its.org/sda7} and the new methodology described here is available in SDA version 7.3, which also includes example files for the systems simulated here.

\begin{acknowledgement}

This research has received funding from the European Union Seventh Framework 
Programme (FP7/2007-2013) under grant agreement number 604102 (HBP Ramp-Up 
Phase), the European Union Horizon 2020 Framework Programme for Research 
and Innovation under grant agreement numbers 720270, 785907 and 945539 (SGA1, SGA2 and SGA3), and the Klaus Tschira Foundation. The authors thank Dr. Stefan Richter for software support, Abraham Muniz-Chicharro for testing of the software, and Dr. Julia Romanowska for initial studies on the protein-surface adsorption systems.

\end{acknowledgement}


\bibliography{references}

\includepdf[pages=-]{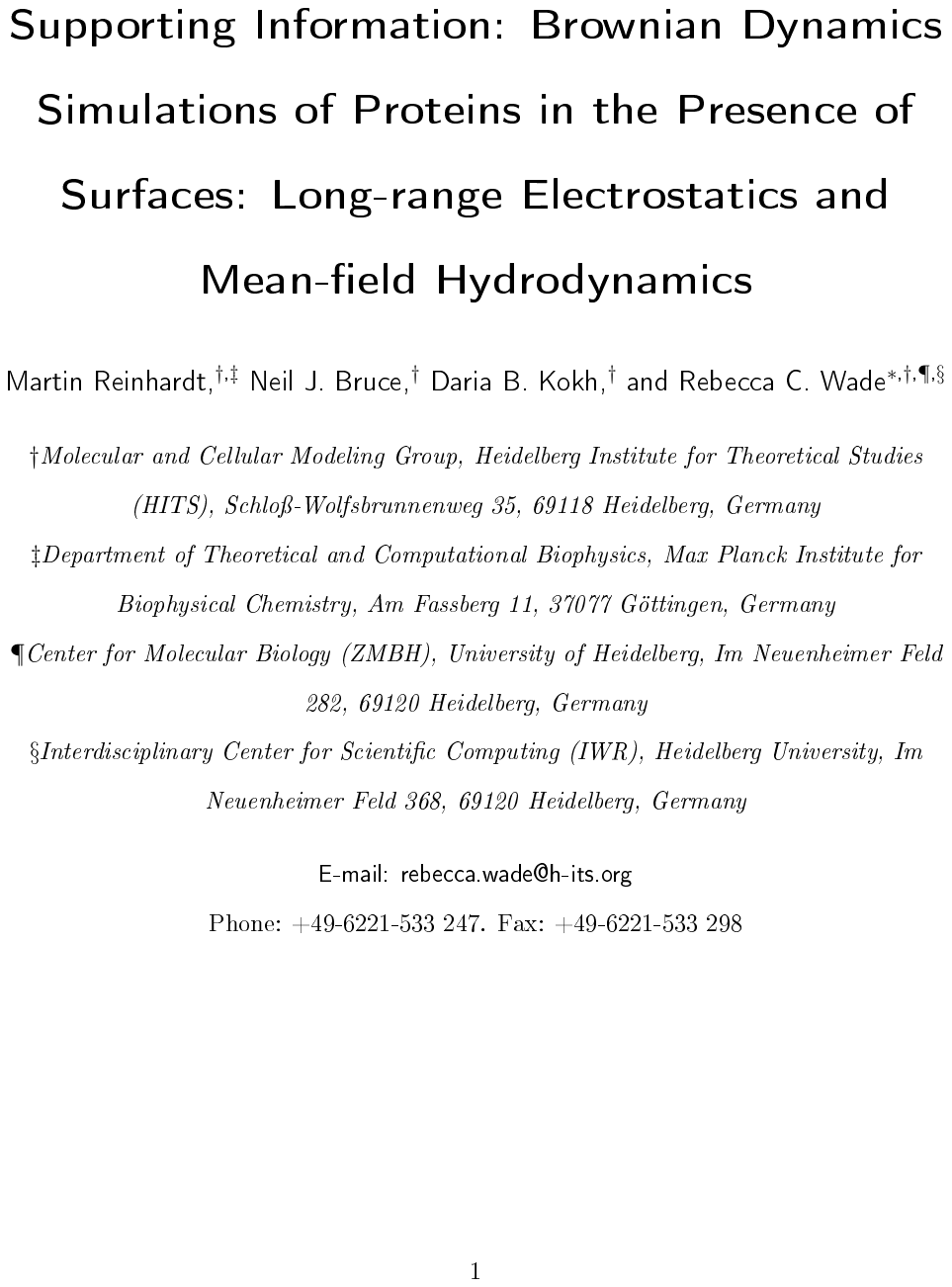}

\end{document}